\documentclass[aps,physrev,preprint,twocolumn,10pt,superscriptaddress]{revtex4-2}
\usepackage{graphicx}% Include figure files
\usepackage{dcolumn}% Align table columns on decimal point
\usepackage{bm}% bold math
\usepackage{amsmath}
\usepackage[colorlinks=true, allcolors=blue]{hyperref}
\usepackage{braket}
\usepackage{lineno}

\newcommand{\posmu}{\ensuremath{\mathrm{\mu^+}}}

\newcommand{\M}{\ensuremath{{\mathrm{M}}}}
\newcommand{\Hy}{\ensuremath{{\mathrm{H}}}}

\begin{document}
%\linenumbers
%\preprint{}

\title{Microwave Spectroscopy of the Muonium $2S_{1/2}-2P_{3/2}$ Fine Structure Interval}

\author{Philipp Blumer}%\email{philipp.blumer@cern.ch}
\affiliation{Institute for Particle Physics and Astrophysics, ETH Zurich, 8093, Switzerland}
\author{Gianluca Janka}
\affiliation{PSI Center for Neutron and Muon Sciences CNM, 5232 Villigen PSI, Switzerland}

\author{Svenja Geissmann}
\affiliation{Institute for Particle Physics and Astrophysics, ETH Zurich, 8093, Switzerland}
\author{Iren Ignatov}
\affiliation{Physics Department, Technion—Israel Institute of Technology, Haifa, 3200003, Israel}
\author{Marcus M\"ahring}
\affiliation{Institute for Particle Physics and Astrophysics, ETH Zurich, 8093, Switzerland}

\author{Zaher Salman}
\affiliation{PSI Center for Neutron and Muon Sciences CNM, 5232 Villigen PSI, Switzerland}
\author{Andreas Suter}
\affiliation{PSI Center for Neutron and Muon Sciences CNM, 5232 Villigen PSI, Switzerland}
\author{Edward Thorpe-Woods}
\affiliation{Institute for Particle Physics and Astrophysics, ETH Zurich, 8093, Switzerland}

\author{Thomas Prokscha}
\affiliation{PSI Center for Neutron and Muon Sciences CNM, 5232 Villigen PSI, Switzerland}

\author{Ben Ohayon}
\affiliation{Physics Department, Technion—Israel Institute of Technology, Haifa, 3200003, Israel}
\author{Paolo Crivelli}\email{crivelli@phys.ethz.ch}
\affiliation{Institute for Particle Physics and Astrophysics, ETH Zurich, 8093, Switzerland}

\collaboration{Mu-MASS Collaboration}

\date{\today}

\begin{abstract}
We report a microwave spectroscopy measurement of the muonium $2S_{1/2}-2P_{3/2}$ fine structure transition, yielding a transition frequency of $9871.0 \pm 7.0~\mathrm{MHz}$, in agreement with state-of-the-art QED predictions within one standard deviation. 
%This represents a fivefold improvement in precision over the only previous measurement. 
In combination with the recent Lamb shift result, the $2P_{1/2}-2P_{3/2}$ splitting is determined, improving the spectroscopic characterization of the $n=2$ manifold in muonium. These results provide a stringent test of bound-state QED in a purely leptonic system and establish a path toward future searches for Lorentz violation and muon-specific new physics.
\end{abstract}

%\keywords{}

\maketitle

\section{Introduction}\label{sec:Intro}

%Recent precision measurements involving the muon have revealed several anomalies that challenge the completeness of the Standard Model. These include the proton charge radius discrepancy observed in muonic hydrogen spectroscopy~\cite{2010_Pohl}, the persistent deviation in the anomalous magnetic moment of the muon ($g-2$)~\cite{2021_mu_g-2}, and indications of lepton flavour universality violation in rare B-meson decays at LHCb~\cite{2022_LHCb}. Together, these findings have prompted the question of whether the muon is simply a heavier analogue of the electron or if it may couple to new physics beyond the Standard Model.

As a purely leptonic atom, Muonium (\M), a hydrogen-like bound state of a positive muon (\posmu) and an electron ($e^-$), is free from hadronic structure effects and finite-size contributions. The absence of nuclear structure makes it a uniquely clean system for high-precision tests of bound-state quantum electrodynamics (QED)\cite{Karshenboim:2005iy,Jentschura:2022xuc,Eides:2023ltp,Patkos:2024lqf,Korobov:2024tlk}. Its relatively long lifetime of $2.2,\mu\mathrm{s}$, in contrast to the much shorter one of positronium, allows for spectroscopic measurements with reduced natural line broadening. Precision spectroscopy of Muonium provides the most accurate determinations of the muon mass and magnetic moment \cite{1999_Liu,2000_Meyer}, thereby serving as an essential input to Standard Model precision tests. Beyond QED studies, Muonium also provides a sensitive probe of new physics, including searches for Lorentz and CPT violation\cite{2014_Gomes,2025_Blumer_FScalc}, couplings to dark sectors~\cite{Frugiuele:2019drl,Stadnik:2022gaa}, and possible new or exotic long-range forces~\cite{Costantino:2019ixl,PhysRevLett.129.239901,Cong:2024qly,Ghosh:2024ctv}.

Since its first observation in 1960~\cite{1960_Hughes}, \M\ spectroscopy has played a central role in precision tests of fundamental physics. Ongoing experimental efforts aim to further improve key transitions: the MuSEUM experiment~\cite{2019_MuSEUM} aims to refine the measurement of the ground-state hyperfine splitting~\cite{1999_Liu}, while within the Mu-MASS collaboration~\cite{2018_Crivelli,2023_Cortinovis} we are pursuing a more accurate determination of the $1S-2S$ transition~\cite{2000_Meyer}.

In contrast, the $n=2$ manifold remains comparatively less explored. In particular, the fine structure (FS) between the $2S_{1/2}$ and $2P_{3/2}$ states, sensitive to relativistic, spin-orbit, and radiative corrections, has been measured only once~\cite{1990_Kettell}, with limited precision. Earlier measurements of the Lamb shift in \M\ faced similar challenges~\cite{1984_Oram,1990_Woodle}, relying on degraders to reduce the muon beam energy. This led to diffuse \M\ distributions, low statistics, and considerable background, ultimately limiting accuracy.

\begin{figure}[h!]
    \centering
     \includegraphics[width=\linewidth, trim=1 1 1 20,clip]{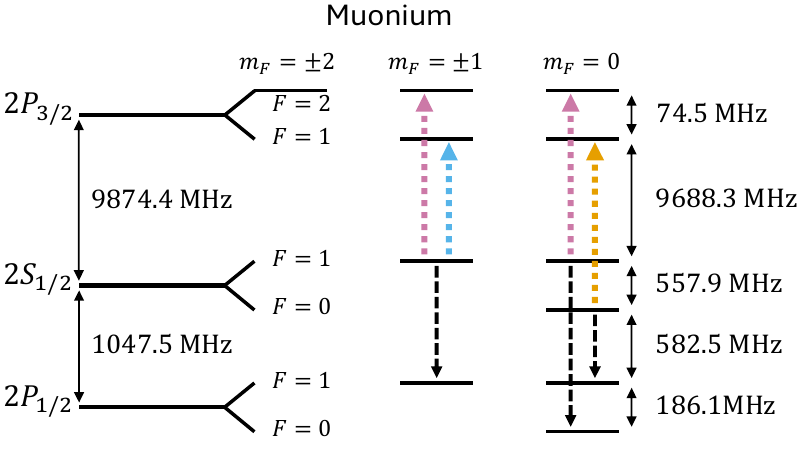}
 \caption{The diagram of the \M\ $n=2$ energy manifold is shown with the allowed electric dipole transitions including the hyperfine structure. The angular momentum $F$ projection along a specified axis is defined with the quantum number $m_F$. The centroid energy of the dashed arrows in black is the so-called Lamb shift. The colored dotted arrows are the fine structure transitions, see Ref. \cite{2025_Blumer_FScalc}.}
    \label{fig:energy_level}
\end{figure}

Recent developments have enabled renewed exploration of the $n=2$ spectrum. The Mu-MASS collaboration at the Paul Scherrer Institute (PSI) performed a high-precision measurement of the $2S_{1/2}-2P_{1/2}$ Lamb shift using an intense metastable $2S$ \M\ beam~\cite{2020_Janka,2022_Ohayon,2022_Janka}. This progress has opened the possibility of accessing the fine structure splitting within the $n=2$ manifold with significantly improved precision. The theoretical prediction for the $2S_{1/2}-2P_{3/2}$ interval, see Fig.~\ref{fig:energy_level}, incorporating higher-order QED corrections, has recently been updated to $9874.367 \pm 0.001\,\mathrm{MHz}$~\cite{2025_Blumer_FScalc}, offering a stringent benchmark for experimental tests.

In this work, we present a dedicated experiment to measure the fine structure of \M. We describe the experimental configuration, data acquisition methods, and spectroscopic analysis that led to the most precise determination of the $2S_{1/2}-2P_{3/2}$ splitting in \M\ to date. Our results improve upon the previous measurement by a factor of five, improving the spectroscopic characterization of the $n=2$ energy levels.

\section{Experimental Methods}
\label{sec:Exp}

We measured the fine structure transition between the $2S_{1/2}$ and $2P_{3/2}$ states in \M\ at the Low Energy Muon (LEM) beamline at the PSI, Switzerland. The experiment uses a single microwave interaction region operated in a zero external magnetic field environment. The general principle involves producing \M\ atoms in the metastable $2S$ state and inducing transitions to the short-lived $2P$ state via resonant microwave radiation. Atoms remaining in the $2S$ state are subsequently quenched by a DC electric field, prompting spontaneous decay to the $1S$ ground state accompanied by the emission of Lyman-$\alpha$ photons ($\lambda = 121\,\mathrm{nm}$), which are then detected.

Due to electric dipole selection rules, direct single-photon decay from the $2S$ to the $1S$ state is forbidden, yielding a radiative lifetime of approximately $122\,\mathrm{ms}$. However, the effective lifetime of the $2S$ state is limited by the muon decay time of $2.2\,\mathrm{\mu s}$. In contrast, the $2P$ state decays rapidly to the $1S$ ground state with a radiative lifetime of approximately $1.6\,\mathrm{ns}$. By scanning the microwave frequency and recording the number of emitted Lyman-$\alpha$ photons, the resonance profile of the transition can be reconstructed.

Before \M\ measurements, the experimental setup was calibrated using atomic hydrogen (\Hy) under analogous conditions. These tests served to validate the Lyman-$\alpha$ detection system and optimize data acquisition and analysis procedures in a controlled environment (see Sec.~\ref{sec:Data}).

\subsection{Muonium Formation at the Low Energy Muon Beamline}

The Swiss Muon Source ($\mathrm{S\mu S}$) at PSI delivers the world's most intense surface muon beam~\cite{2008_LEM}. Muons are produced from $590\,\mathrm{MeV}$ protons impinging on a graphite target, generating pions that decay at rest near the surface into fully polarized \posmu\ via weak decay~\cite{2024_muSR}. At the LEM beamline, a fraction of these muons is moderated to $\sim15\,\mathrm{eV}$ using a thin solid noble gas layer~\cite{1994_Morenzoni,2001_Prokscha,2004_Morenzoni}. They are then re-accelerated electrostatically to kinetic energies between $1$ and $20\,\mathrm{keV}$.

\begin{figure*}[t!]
    \centering
    \includegraphics[width=0.9\textwidth, trim=1 1 1 1, clip]{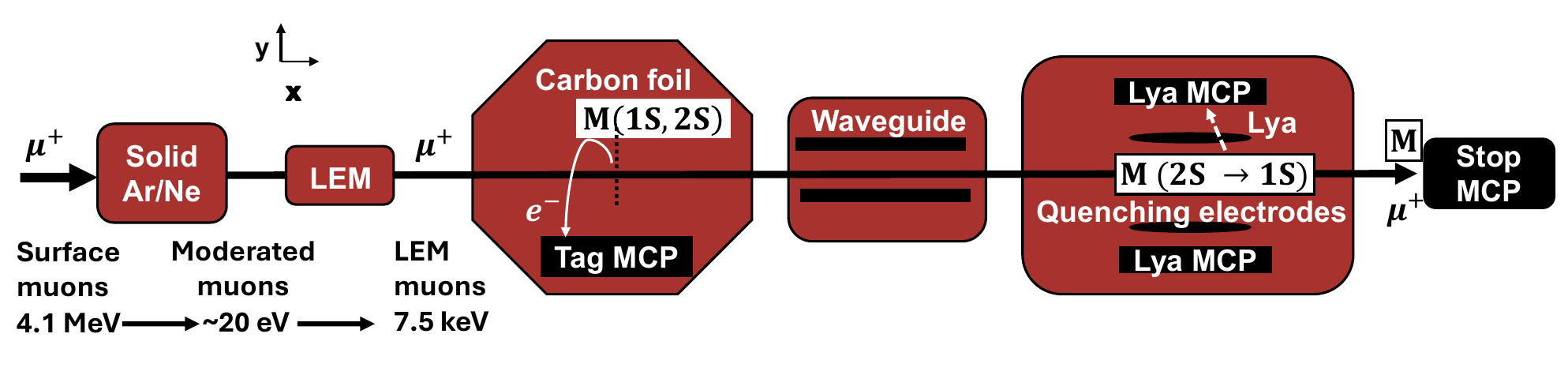}
    \caption{Scheme of the fine structure experimental setup. The transition rate is determined from coincidence signals between the tagging, Lyman-$\alpha$, and stopping MCP detectors. After forming neutral atoms via interaction with an ultrathin carbon foil, a WR90 waveguide induces the frequency-dependent $2S_{1/2}-2P_{3/2}$ energy transition.}
    \label{fig:fs_scheme}
\end{figure*}

Electrostatic beam optics and an $\vec{E}\times\vec{B}$ spin rotator steer and purify the beam, effectively removing unwanted backgrounds such as protons~\cite{2012_Salman}. The muons are then directed to the experimental fine structure apparatus; see Fig.~\ref{fig:fs_scheme}.

When \posmu\ are not available, a proton beam can be generated by thermionic emission of electrons from a heated tantalum filament, followed by proton generation through ionization of water frozen on the moderator ~\cite{2024_Janka_LEM}. These protons are then extracted electrostatically and guided through the apparatus after appropriate spin rotator and optics adjustments, serving as proxies for system alignment and commissioning.

In both operational modes, positively charged particles are focused onto a $2\,\mathrm{nm}$ ultrathin carbon foil, previously characterized by the LEM facility~\cite{2024_Janka_LEM}. Passage through the foil results in kinetic energy loss and secondary electron emission. Electrons ejected near the foil surface are collected by a microchannel plate detector (\textit{Tag MCP}) due to a positively biased front plate, which provides a start signal for time-of-flight measurements~\cite{1995_CFoil}.

A fraction of particles exiting the foil undergoes electron capture, forming neutral atoms, either \Hy~\cite{2015_Allegrini,2016_Allegrini} or \M. For incident \posmu\ with $7.5\,\mathrm{keV}$ energy, approximately $50\%$ form neutral \M~\cite{2024_Janka_LEM}, and among these, $11\pm4\%$ populate the excited $2S$ state~\cite{2020_Janka}.

%\begin{figure}[htb]
  %  \centering
%    \includegraphics[width=\linewidth, trim=1 1 1 1,clip]{scheme_fs_v2.pdf}
 %   \caption{Scheme of the fine structure experimental setup. The transition rate is determined from coincidence signals between the tagging, Lyma%n-$\alpha$, and stopping MCP detectors. After forming neutral atoms via interaction with an ultrathin carbon foil, a WR90 waveguide induces the %   \label{fig:fs_scheme}
%\end{figure}

\subsection{Frequency-Dependent State Detection}

The excitation of the $2S_{1/2}-2P_{3/2}$ transition is driven by a modified WR90 microwave waveguide operating in the $8$–$11\,\mathrm{GHz}$ frequency range. The microwave field is coupled into the interaction region via a frequency generator (\textsc{AnaPico APSIN26G}), while the output power is continuously monitored by a power sensor (\textsc{Keysight U2002A}). The waveguide has an internal aperture of $18\,\mathrm{mm} \times 10\,\mathrm{mm}$ aligned with the neutral atom beam trajectory and is enclosed by a copper grid to suppress electromagnetic field leakage.

Both the input frequency and the measured output power are remotely controlled and recorded via the slow control system. This configuration allows for active stabilization of the electric field amplitude $V_0$ throughout the frequency scan, by compensating for frequency-dependent transmission losses through the waveguide. When the field is resonant with the $2S_{1/2}-2P_{3/2}$ transition, a subset of atoms is promoted to the $2P$ state and decays within $1.6\,\mathrm{ns}$, emitting a Lyman-$\alpha$ photon. The excitation rate is thus encoded in the depletion of the surviving $2S$ population and the correlated emission of ultraviolet photons.

\subsection{Lyman-$\alpha$ Detection System}

Following the interaction region, neutral atoms traverse a detection system consisting of two CsI-coated MCP detectors (\textit{LyA MCPs}), separated by a static electric field of $250\,\mathrm{V/cm}$. This field mixes the residual $2S$ population with the $2P$ states, accelerating their relaxation to the $1S$ ground state. The resulting Lyman-$\alpha$ photons interact with the CsI coating, which enhances the quantum efficiency up to $50\%$ for photon-to-electron conversion~\cite{2005_MCP_QE,2017_MCP_UV}.

At the downstream end of the beamline, the particles are intercepted by a third MCP detector (\textit{Stop MCP}), which serves as a trigger for time-of-flight measurements. 
%The detection window is defined as the $13\,\mathrm{\mu s}$ preceding the trigger, corresponding to approximately six muon lifetimes, and allowing multiple hits in the \textit{LyA MCP} within this interval.
A typical run accumulates $\sim10^5$ events in $15\,\mathrm{min}$. Coincidence analysis across the \textit{Tag MCP}, \textit{LyA MCPs}, and \textit{Stop MCP} enables time-correlated reconstruction of each excitation-decay sequence, processed during offline analysis.

\subsection{Measurement Campaign}

During the PSI accelerator service, the apparatus was initially commissioned using \Hy\ atoms formed from protons at the LEM beamline. High-voltage supplies and microwave infrastructure were integrated into the LEM facility’s slow control environment, enabling fully remote operation. During commissioning, it was observed that only one \textit{LyA MCP} could be operated at a time because of signal crosstalk between detectors.

Two reference frequency microwave scans of the \Hy\ fine structure transition were performed using an argon ($\mathrm{Ar}$) moderator.
The microwave field strength was $20\,\mathrm{V/cm}$ and the microwave propagation direction was reversed between scans (A/B configurations) to verify field uniformity and to control for possible Doppler shifts arising from a misalignment between the atomic beam and
the travelling-wave microwave field.

Five \M\ lineshape scans were completed following the \Hy\ tests over six consecutive days. The first used an $\mathrm{Ar}$ moderator, while the remaining scans employed a more efficient neon ($\mathrm{Ne}$) moderator, increasing the \posmu\ rate by a factor of $\sim1.5$. The electric field amplitude of the waveguide $V_0$ and the propagation direction were systematically varied, as summarized in Table~\ref{tab:statistic-summary}. 
These changes allowed for control and verification of the microwave field characteristics and the assessment of Doppler shifts as a result of atomic motion relative to the RF propagation direction. By reversing the microwave direction, the Doppler effect changes sign, enabling its isolation from other systematic effects.

\section{Data Analysis}
\label{sec:Data}

\begin{figure}[htb!]
    \centering
    \includegraphics[width=\linewidth]{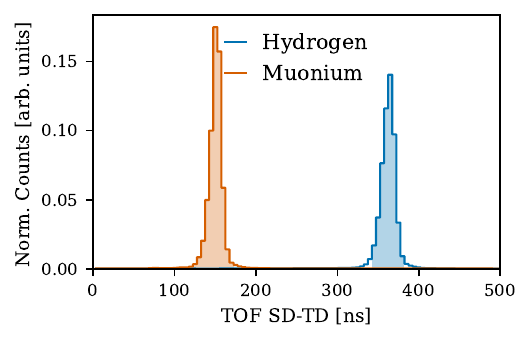}
    \caption{The first cut applied to the datasets involves the tagging (TD) and stopping (SD) detectors, resulting in an energy selection. The TOF difference between \Hy\ (blue) and \M\ (orange) reflects the mass difference, with \M\ traveling faster for the same kinetic energy. Events with a relative TOF outside of the shaded regions are excluded.}
    \label{fig:norm_count}
\end{figure}

The analysis begins with processing the time-stamped signals from the MCP detectors, following the methodology established in the \M\ Lamb shift measurement~\cite{2022_Ohayon,2022_Janka}. The normalization signal $S_\mathrm{norm}$ is defined as the number of events with a coincidence between the \textit{Tag} and \textit{Stop MCPs} within a predefined time window. Because of the known separation between these detectors and the expected kinetic energy of particles after the carbon foil, this selection effectively applies an energy cut, independent of the waveguide frequency, see Fig.~\ref{fig:norm_count}.

Time-of-flight (TOF) signals between the \textit{LyA} and the \textit{Tag/Stop MCPs}, respectively, are then used to isolate the $2S$ quenching signal. These are verified with frequency-dependent microwave \textit{On/Off} measurements, producing the triple-coincidence signal $S_\gamma$ from Lyman-$\alpha$ photons. The final normalized signal as a function of microwave frequency $f$ is defined as:
\begin{equation}
    S(f)=\frac{S_\gamma(f)}{S_\mathrm{norm}},
\end{equation}
which compensates for long-term beam intensity drifts.

The tagging rate, defined as $S_\mathrm{norm}/\Delta t$, was continuously recorded to monitor and optimize data collection efficiency. For \M, this rate was additionally normalized to the PSI accelerator current (in $\mathrm{mA}$)\footnote{The typical proton current during normal accelerator operation is around 2 $\mathrm{mA}$.}  to account for fluctuations in muon production. This allowed for real-time tracking of neutral atom formation, helping to optimize moderator regrowth cycles and detect changes in beam or detector performance. Across all \M\ scans, the average tagging rate varied between $15.5$ and $24.5\,\mathrm{Hz/mA}$, as summarized in Tab.~\ref{tab:statistic-summary}. The \textit{Tag MCP} has a measured efficiency of approximately $65\%$, setting the effective upper limit on detected \M\ rates.

\begin{table*}[htb!]
    \centering
    \begin{tabular}{c|cc|cc||cc|cc|cc|cc|cc}
    \hline
        & \multicolumn{4}{c||}{\textbf{Hydrogen}} & \multicolumn{10}{c}{\textbf{Muonium}}\\
    \hline
        & \multicolumn{2}{c|}{$\Hy\ 1$} & \multicolumn{2}{c||}{$\Hy\ 2$} & \multicolumn{2}{c|}{$\M\ 1$} & \multicolumn{2}{c|}{$\M\ 2$} & \multicolumn{2}{c|}{$\M\ 3$} & \multicolumn{2}{c|}{$\M\ 4$} & \multicolumn{2}{c}{$\M\ 5$} \\
    Frequency & $S_\mathrm{\gamma}$ & $S_\mathrm{norm}$ & $S_\mathrm{\gamma}$ & $S_\mathrm{norm}$ & $S_\mathrm{\gamma}$ & $S_\mathrm{norm}$ & $S_\mathrm{\gamma}$ & $S_\mathrm{norm}$ & $S_\mathrm{\gamma}$ & $S_\mathrm{norm}$ & $S_\mathrm{\gamma}$ & $S_\mathrm{norm}$ & $S_\mathrm{\gamma}$ & $S_\mathrm{norm}$ \\
    $\mathrm{MHz}$  &       & ($10^6$)  &       & ($10^6$)  &       & ($10^6$)  &       & ($10^6$)  &       & ($10^6$)  &       & ($10^6$)  &       & ($10^6$) \\
    \hline
    $8000$  & $184$   & $1.31$  & $177$   & $0.71$  & $133$   & $0.19$  & $127$   & $0.23$  & $195$   & $0.35$  & $136$   & $0.22$  & $212$   & $0.29$  \\
    $9000$  &       &       &       &       & $142$   & $0.22$  & $162$   & $0.25$  & $199$   & $0.33$  & $165$   & $0.25$  & $231$   & $0.35$  \\ 
    $9300$  & $218$   & $1.35$  & $212$   & $0.90$  &       &       &       &       &       &       &       &       &       &       \\
    $9550$  &       &       &       &       & $80$    & $0.17$  & $106$   & $0.22$  & $205$   & $0.38$  & $134$   & $0.21$  & $181$   & $0.32$  \\
    $9600$  & $215$   & $1.34$  & $186$   & $0.87$  &       &       &       &       &       &       &       &       &       &       \\
    $9675$  &       &       &       &       & $49$    & $0.17$  & $63$    & $0.22$  & $142$   & $0.36$  & $101$   & $0.22$  & $110$   & $0.31$  \\
    $9775$  &       &       &       &       & $72$    & $0.20$  & $71$    & $0.25$  & $116$   & $0.35$  & $115$   & $0.24$  & $91$    & $0.35$  \\
    $9800$  & $146$   & $1.32$  & $123$   & $0.70$  &       &       &       &       &       &       &       &       &       &       \\
    $9850$  &       &       &       &       & $79$    & $0.20$  & $57$    & $0.17$  & $132$   & $0.38$  & $98$    & $0.23$  & $141$   & $0.32$  \\
    $9875$  & $96$    & $1.33$  & $69$    & $0.71$  &       &       &       &       &       &       &       &       &       &       \\
    $9925$  & $101$   & $1.32$  & $97$    & $0.71$  &       &       &       &       &       &       &       &       &       &       \\
    $10000$ &       &       &       &       & $102$   & $0.20$  & $104$   & $0.22$  & $207$   & $0.35$  & $135$   & $0.21$  & $184$   & $0.29$  \\
    $10025$ & $165$   & $1.32$  & $120$   & $0.71$  &       &       &       &       &       &       &       &       &       &       \\
    $10100$ & $198$   & $1.33$  & $196$   & $0.87$  &       &       &       &       &       &       &       &       &       &       \\
    $10175$ &       &       &       &       & $104$   & $0.17$  & $82$    & $0.16$  & $187$   & $0.35$  & $105$   & $0.21$  & $156$   & $0.32$  \\
    $10250$ &       &       &       &       & $107$   & $0.20$  & $128$   & $0.25$  & $189$   & $0.38$  & $141$   & $0.23$  & $171$   & $0.35$  \\
    $10400$ & $204$   & $1.34$  & $210$   & $0.90$  & $109$   & $0.20$  & $124$   & $0.22$  & $234$   & $0.38$  & $158$   & $0.24$  & $206$   & $0.35$  \\
    $10800$ &       &       &       &       & $127$   &$ 0.20$  & $145$   & $0.22$  & $195$   & $0.33$  & $147$   & $0.21$  & $172$   & $0.29 $ \\
    \hline%\hline
    Time    &  \multicolumn{2}{c|}{$7.5\,\mathrm{h}$}   &  \multicolumn{2}{c||}{$6.7\,\mathrm{h}$}   &  \multicolumn{2}{c|}{$19.7\,\mathrm{h}$}  &  \multicolumn{2}{c|}{$16.8\,\mathrm{h}$}  &  \multicolumn{2}{c|}{$36.6\,\mathrm{h}$}  &  \multicolumn{2}{c|}{$19.4\,\mathrm{h}$}  &  \multicolumn{2}{c}{$21.6\,\mathrm{h}$} \\
    \hline
    Rate &  \multicolumn{2}{c|}{$440\,\mathrm{Hz}$}  &  \multicolumn{2}{c||}{$293\,\mathrm{Hz}$}  &  \multicolumn{2}{c|}{$15.5\,\mathrm{Hz/mA}$}   &  \multicolumn{2}{c|}{$21.3\,\mathrm{Hz/mA}$}   &  \multicolumn{2}{c|}{$19.4\,\mathrm{Hz/mA}$}   &  \multicolumn{2}{c|}{$20.1\,\mathrm{Hz/mA}$}   &  \multicolumn{2}{c}{$24.5\,\mathrm{Hz/mA}$} \\
    \hline  
    \end{tabular}
    \caption{Statistics collected during the beamtime for the \Hy\ and \M\ fine structure measurements. Each entry lists the number of Lyman-$\alpha$ photon events $S_\gamma$ and normalization events $S_\mathrm{norm}$ (in millions) for the respective frequency points. The rate is the triple-coincidence signal between \textit{LyA} and the \textit{Tag/Stop MCPs}.}
    \label{tab:statistic-summary}
\end{table*}

The microwave output power was continuously recorded at $5\,\mathrm{s}$ intervals to monitor experimental stability. Only data with relative power fluctuations within $\pm10\%$ of the nominal output were retained for each frequency scan point. This corresponds to variations of approximately $\pm5\,\mathrm{V/cm}$ in the waveguide electric field amplitude $V_0$. Datasets exceeding this threshold were excluded from the analysis to avoid introducing systematic distortions to the lineshape.

\begin{figure}[htb!]
    \centering
    \includegraphics[width=\linewidth]{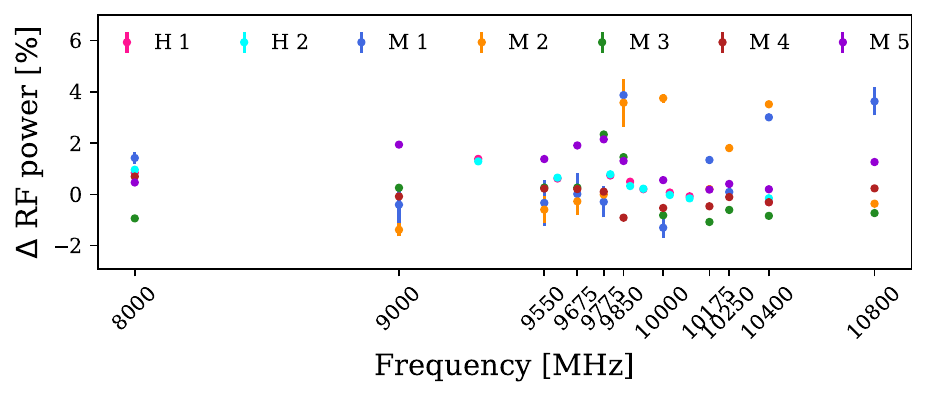}
    \caption{The relative deviation of the measured output power is less than $4\pm 1\%$, confirming microwave power control during the measurements.}
    \label{fig:wg_avg}
\end{figure}

This stability criterion led to the rejection of $10\%$ of the $\M\ 2$ dataset and $7.5\%$ of $\M\ 3$ (see Table \ref{tab:lineshape_scans}), due to disruptions in the communication between the slow control system and the recording device. Across all accepted scans, the weighted average of frequency-dependent microwave power remained within $4\pm 1\%$ of the target, confirming the consistency of the applied threshold, see Fig.~\ref{fig:wg_avg}.

\subsection{Simulation-Based Data Fitting}

\begin{figure*}[htb!]
    \centering
   % \begin{subfigure}[t]{0.48\linewidth}
      %  \centering
        \includegraphics[width=0.45\textwidth]{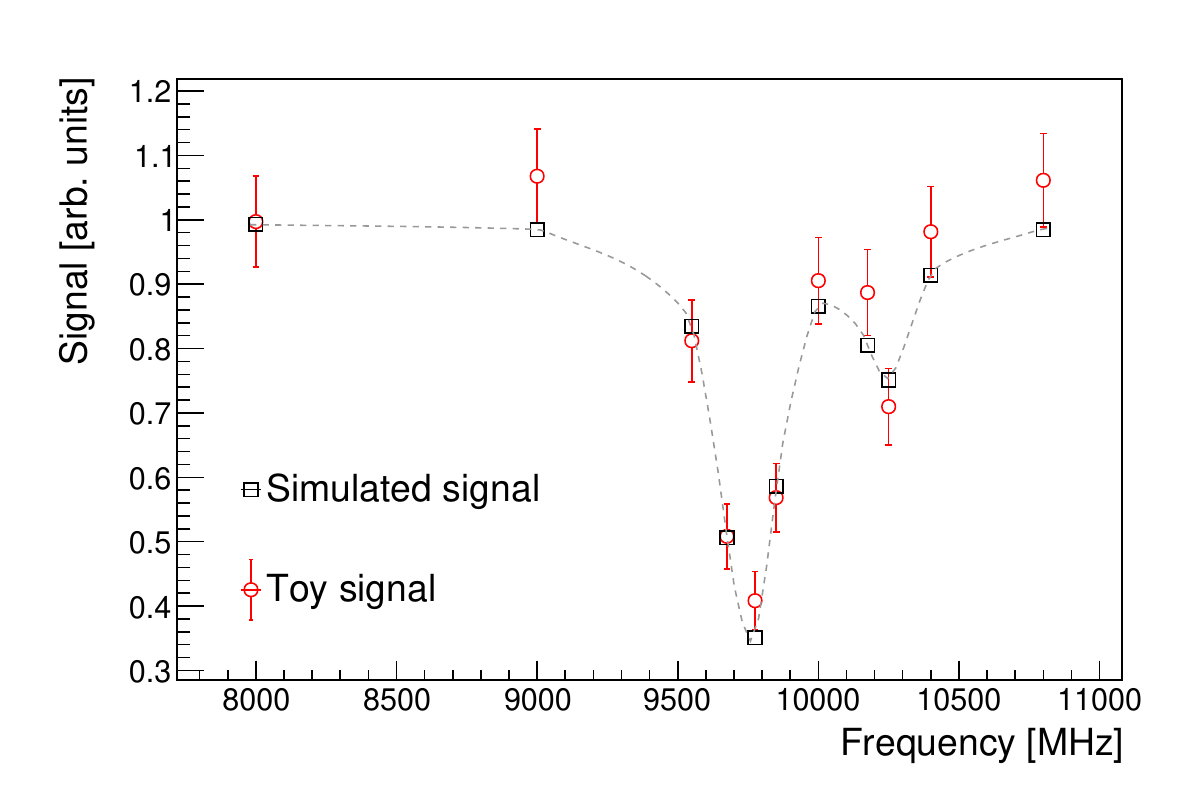}
        \includegraphics[width=0.45\textwidth]{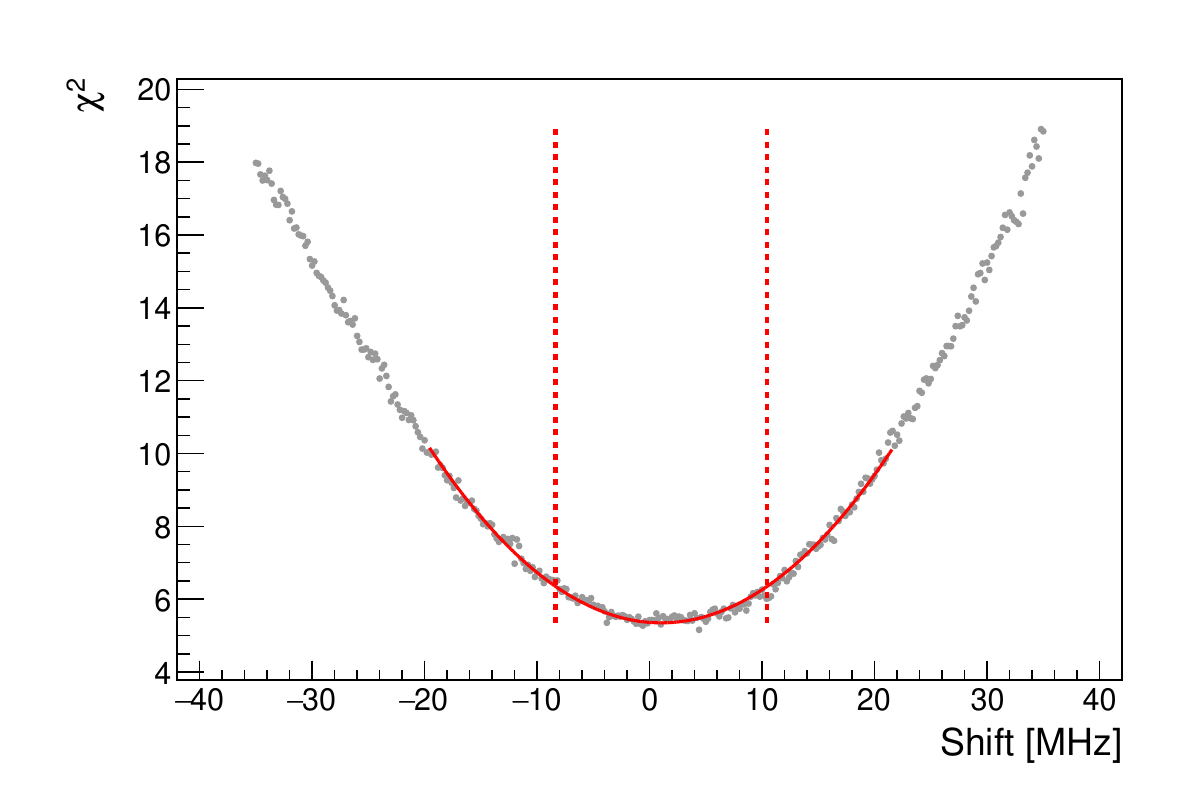}
        \caption{Validation of the fitting procedure using simulated toy data and $\chi^2$ minimization. Left: simulated \M\ transition probabilities $p(f)$ (black squares) and corresponding toy data (red circles) used to validate the fitting procedure. The dashed gray line indicates the underlying lineshape for visualization only. Right:$\chi^2$ distribution from the least-squares fit to the simulated toy data. The dashed lines indicate the $1\sigma$ confidence interval as defined in Eq.~\ref{eq:chi2_err}}
       % \label{fig:toy_lineshape}
    %\end{subfigure}
    %\hfill
    %\begin{subfigure}[t]{0.48\linewidth}
     %   \centering
       % \includegraphics[width=\linewidth]{chi2_minimization_M_toy_Ne60.pdf}
       % \caption{$\chi^2$ distribution from the least-squares fit to the simulated toy data in Fig.~\ref{fig:toy_lineshape}. The dashed lines indicate the $1\sigma$ confidence interval as defined in Eq.~\ref{eq:chi2_err}.}
       % \label{fig:toy_chi2}
    %\end{subfigure}
    %\caption{Validation of the fitting procedure using simulated toy data and $\chi^2$ minimization.}
    \label{fig:simulation}
\end{figure*}

Our data analysis procedure relies on detailed Monte Carlo simulations of the LEM beamline~\cite{2012_musrSIM,2015_LEMsim}, based on the \textsc{Geant4} framework~\cite{2003_Geant4} and continuously refined by the Mu-Mass collaboration~\cite{2022_Ohayon,2022_Janka,2024_CortinovisPhD}. High-statistics simulations of the \Hy\ and \M\ atomic beams include realistic energy and momentum distributions downstream of the carbon foil and the applied high-voltage settings. Atoms are initialized in the $2S_{1/2}$ hyperfine sublevels with populations $(0.25, 0.25, 0.5)$ for $(F=0, m_F=0)$, $(F=1, m_F=0)$, and $(F=1, m_F=\pm1)$, respectively.

Atomic trajectories through the microwave interaction region are computed using field maps from \textsc{CST 3D-EM} (a commercial 3D electromagnetic field solver), incorporating standing-wave patterns and reflections in the modified WR90 waveguide. The time-dependent electric field in the simulation is described by:
\begin{equation}
\label{eq:sim_Efield}
    \vec{E}(\vec{r},t) = 
    A \Bigl[
        \mathrm{Re}\bigl(\vec{E}(\vec{r})\bigr)\cos(\omega t + \delta)
        + \mathrm{Im}\bigl(\vec{E}(\vec{r})\bigr)\sin(\omega t + \delta)
    \Bigr],
\end{equation}
where $A = \sqrt{P_\text{WG}^\text{measured} / P_\text{sim}}$ scales the simulated waveguide power (as obtained from the CST 3D-EM field maps) to the measured power, $\omega=2\pi f$ is the angular frequency, and $\delta$ is a random phase.

RF-driven transition dynamics are computed using an adaptive Runge-Kutta integrator~\cite{1992_RungeKutta} applied to two- or three-level optical Bloch equations, depending on the state configuration (see Appendix~\ref{app:Bloch}). For each atom, a random number $r \in [0,1]$ is drawn and compared with the computed excitation probability; atoms that exceed this threshold are considered to undergo a $2S$ to $2P$ transition and emit a Lyman-$\alpha$ photon. The resulting transition probabilities $p(f)$ are stored in high-statistics \textsc{Root} histograms~\cite{1997_ROOT} for each frequency point, providing the basis for fitting, as illustrated in Fig.~\ref{fig:simulation}.

The measured lineshape is fitted using a linear model based on the simulated transition probabilities:
\begin{equation}
\label{eq:fit}
    \lambda(f;a,b) = a \cdot p(f) + b,
\end{equation}
where $a$ is a scaling factor and $b$ represents a flat background offset. The optimal parameters $(a,b)$ are obtained by minimizing the chi-squared function:
\begin{equation}
\label{eq:chi2_min}
    \chi^2 = \sum_{i=1}^{N} \frac{ \bigl(S_i - \lambda_i \bigr)^2 }{\sigma_i^2 },
\end{equation}
where $S_i$ is the measured signal at frequency $f_i$ and "$\sigma_i$ denotes the statistical uncertainty of the measured signal, taken as 
$\sigma_i=\sqrt{S_i}$, assuming Poisson statistics~\cite{1998_CowanStat}.

The $\chi^2$ is evaluated for frequency-shifted histograms to extract the resonance location. Around the minimum, the $\chi^2$ curve is well-approximated by a parabola, and the $1\sigma$ statistical uncertainty is determined by:
\begin{equation}
\label{eq:chi2_err}
    \chi^2_\sigma = \chi^2_\mathrm{min} + 1.
\end{equation}

Before fitting the \M\ data, the complete analysis chain was validated with toy datasets generated from Gaussian fluctuations around the simulated transition probabilities $p(f)$, as shown in Fig.~\ref{fig:simulation}. The procedure was first benchmarked with the \Hy\ datasets, where the extracted $2S_{1/2}-2P_{3/2}$ resonance is consistent with known \Hy\ fine structure results~\cite{1994_Hagley_Hydrogen_FS}. Only after this validation, and once the analysis was finalized, it was applied to the muonium data, ensuring that no adjustments were made after the fit was performed. For \M, the larger asymmetries in the lineshape required an expanded frequency scan of $\pm 35\,\mathrm{MHz}$ with $0.2\,\mathrm{MHz}$ binning to achieve robust resonance extraction.

\section{Results and Discussion}
\label{sec:Results}

\begin{table*}[htb!]
    \centering
    \begin{tabular}{c||c c c c c c c}
         & Moderator   & Waveguide $V_0$ & RF direction & Shift [MHz] & $\chi_\text{min}^2$ / ndof & $a$ [$10^{-3}$] & $b$ [$10^{-3}$]\\
        \hline \hline
        \Hy\ 1 & $\mathrm{Ar}$ & $20\,\mathrm{V/cm}$ & A & $14.8\pm 9.5$ & $9.2 / 7$ & $0.19\pm 0.02$ & $-0.03\pm 0.01$\\
        \Hy\ 2 & $\mathrm{Ar}$ & $20\,\mathrm{V/cm}$ & B & $2.1\pm 9.1$ & $2.6 / 7$ & $0.3\pm 0.03$ & $0.08\pm 0.03$\\
        \hline
        \M\ 1  & $\mathrm{Ar}$ & $60\,\mathrm{V/cm}$ & B & $-34.0\pm 17.8$ & $9.8 / 9$ & $0.53\pm 0.07$ & $0.10\pm 0.05$\\
        \M\ 2  & $\mathrm{Ne}$ & $60\,\mathrm{V/cm}$ & B & $-15.5\pm 14.4$ & $9.1 / 9$ & $0.55\pm 0.06$ &  $0.05\pm 0.05$\\
        \M\ 3  & $\mathrm{Ne}$ & $30\,\mathrm{V/cm}$ & B & $22.9\pm 12.8$ & $6.2 / 9$ & $1.0\pm 0.1$ & $-0.4\pm 0.1$\\
        \M\ 4  & $\mathrm{Ne}$ & $20\,\mathrm{V/cm}$ & B & $16.6\pm 27.0$ & $11.5 / 9$ & $1.6\pm 0.3$ & $-0.9\pm 0.3$\\
        \M\ 5  & $\mathrm{Ne}$ & $30\,\mathrm{V/cm}$ & A & $-6.0\pm 9.1$ & $7.7 / 9$ & $1.3\pm 0.1$ & $-0.6\pm 0.1$\\
        \hline
    \end{tabular}
    \caption{Summary of \Hy\ and \M\ lineshape scans performed at the LEM beamline. Variations include the moderator material ($\mathrm{Ar}$, $\mathrm{Ne}$), electric field amplitude, and microwave propagation direction (A/B).}
    \label{tab:lineshape_scans}
\end{table*}

\begin{figure*}[htb!]
    \centering
        \includegraphics[width=0.45\textwidth]{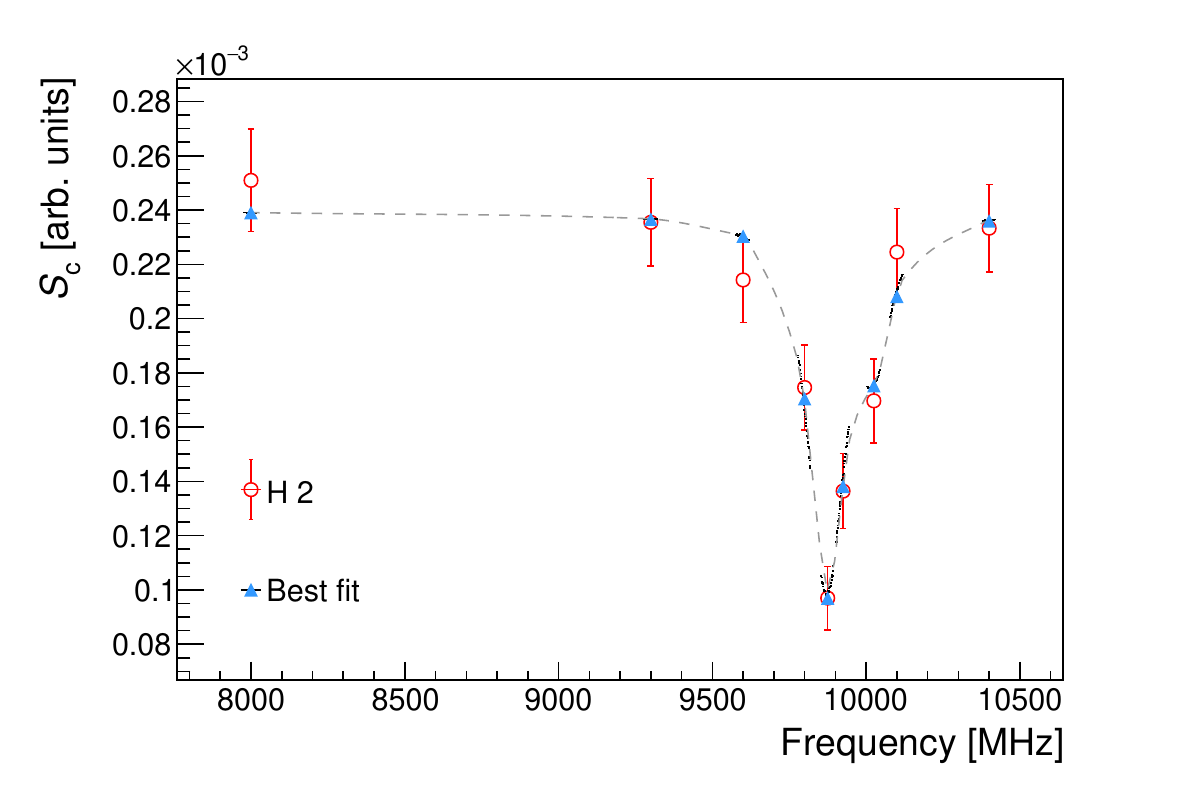}
        \includegraphics[width=0.45\textwidth]{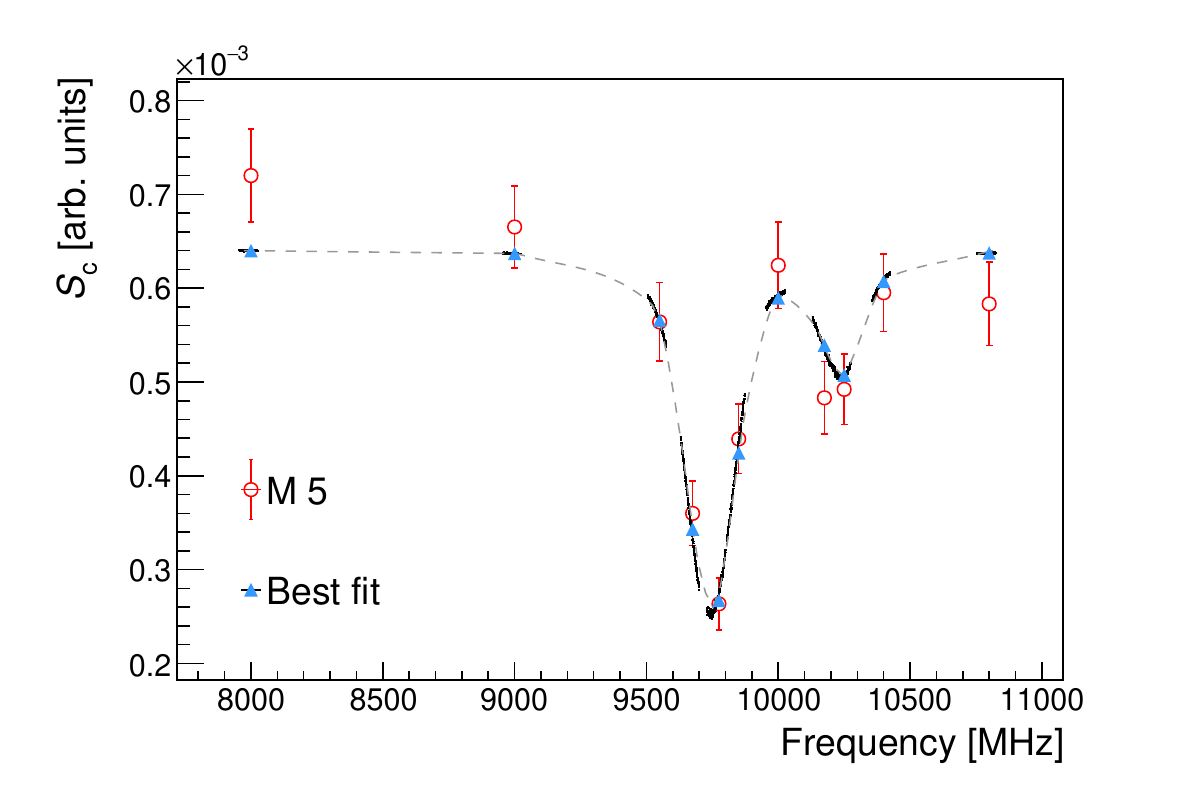}
        \caption{Comparison of $\Hy\ 2$ and $\M\ 5$ resonance scans. The dashed gray lines indicate the underlying lineshapes for visualization only. Left: Hydrogen scan $\Hy\ 2$ in red circles with the corresponding best fit from the simulation represented by blue triangles: $f_\mathrm{\Hy 2} = 2.1\pm 9.1\,\mathrm{MHz}$, $\chi_\mathrm{min,\Hy 2}^2 = 2.6$. Right: Muonium scan $\M\ 5$ in red circles with the corresponding best fit from the simulation represented by blue triangles: $f_\mathrm{\M 5} = -6.0\pm 9.1\,\mathrm{MHz}$, $\chi_\mathrm{min,\M 5}^2 = 7.7$. }
    \label{fig:HM_compare}
\end{figure*}

\subsection{Hydrogen Validation Measurements}

Independent \Hy\ fine structure scans were performed with a waveguide electric field of $V_0 = 20\,\mathrm{V/cm}$ and opposite RF propagation directions (A/B), as summarized in Tab.~\ref{tab:lineshape_scans}, to systematically validate the experimental setup. The frequency shifts obtained from the fit, relative to the known experimental and theoretical values for the hydrogen fine-structure transition, were:
\begin{align*}
    \Hy\ 1 &: 14.8 \pm 9.5\,\mathrm{MHz},\\
    \Hy\ 2 &: 2.1 \pm 9.1\,\mathrm{MHz}.
\end{align*}
The fits use two free parameters with nine frequency points, resulting in $\nu = 7$ degrees of freedom. The corresponding minimal chi-squared values are $\chi^2_\mathrm{min} = 9.2$ ($\Hy\ 1$) and $2.6$ ($\Hy\ 2$). These values are within the expected statistical range for the given number of degrees of freedom, indicating that the model describes the data adequately. Fig.~\ref{fig:HM_compare} on the left shows the measured lineshape for $\Hy\ 2$ along with the best-fit simulation. 

The combined weighted average, calculated as
\begin{equation}
\label{eq:weighted_avg}
    \hat{\mu} = 
    \frac{\sum_i y_i / \sigma_i^2}{\sum_i 1 / \sigma_i^2},
\end{equation}
yields $\hat{\mu}_\Hy = 8.1 \pm 6.6\,\mathrm{MHz}$. Subtracting this shift from the theoretical prediction gives an experimental resonance frequency of $9903.1 \pm 6.6$MHz, in agreement with the theoretical value 
$9911.2093\pm 0.0001\,\mathrm{MHz}$~\cite{2016_Horbatsch_Htabulation} and the most precise measurement to date 
$9911.200 \pm 0.012\,\mathrm{MHz}$~\cite{1994_Hagley_Hydrogen_FS}.  
These tests confirmed the stability and accuracy of the setup and data analysis pipeline, enabling the transition to \M\ spectroscopy.

\subsection{Muonium Fine Structure Measurement}
\label{subsec:M_results}

The \M\ fine structure was probed at eleven frequency points under five different experimental conditions (Tab.~\ref{tab:lineshape_scans}), giving $\nu=9$ degrees of freedom per fit. The \M\ rate increased by a factor $1.5$ after switching the moderator from $\mathrm{Ar}$ to $\mathrm{Ne}$, resulting in shorter yet more precise measurements, e.g., compare $\M\ 1$ to $\M\ 2$. However, measurement $\M\ 3$ was affected by a reduced \posmu\ rate and connectivity issues between the LEM slow control and the microwave system. Compared to $\M\ 4$, the waveguide field direction was switched in $\M\ 5$ as a first attempt to determine the Doppler shift. The measured lineshape of the $\M\ 5$ scan is shown in Fig.~\ref{fig:HM_compare} on the right. The individual frequency shifts and corresponding $\chi_\mathrm{min}^2$ values are consistent with expectations. 

\begin{figure}[htb!]
    \centering
    \includegraphics[width=\linewidth]{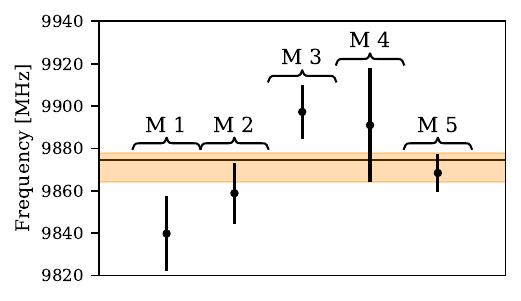}
    \caption{Independent \M\ fine structure measurements. The black line marks the theoretical prediction, and the orange band indicates the one-sigma confidence interval of the weighted average.}
    \label{fig:results}
\end{figure}

Figure~\ref{fig:results} shows the independent results compared to their weighted mean. The combined statistical weighted average yields
$$
    \hat{\mu}_\M = -3.4 \pm 6.0\,\mathrm{MHz},
$$
with a p-value of $0.71$, suggesting the data to be consistent. The systematic uncertainties are summarized in Tab.~\ref{tab:M_results}. The dominant systematic arises from a potential waveguide–beam misalignment, which introduces a first-order Doppler uncertainty of $3.2\,\mathrm{MHz}$ as determined by \textsc{Geant4}-based simulations assuming a $30\,\mathrm{mrad}$ mechanical tolerance. Additional contributions come from RF field map variations ($1.2\,\mathrm{MHz}$), monitored RF power stability ($0.8\,\mathrm{MHz}$), and the second-order Doppler effect ($0.6\,\mathrm{MHz}$). The impact of velocity distribution uncertainties is negligible ($<0.01\,\mathrm{MHz}$), and population of higher excited states ($n\ge3$) can be excluded for this transition range.

The final result for the $2S_{1/2}-2P_{3/2}$ transition is:
$$
    \nu_\mathrm{FS}^\M = 9871.0 \pm 7.0\,\mathrm{MHz},
$$
where the uncertainty includes both statistical and systematic contributions, agreeing with the theoretical prediction of $9874.367 \pm 0.001\,\mathrm{MHz}$~\cite{2025_Blumer_FScalc} and improving the only previous measurement by a factor of five~\cite{1990_Kettell}.  

Combining this result with the most recent Lamb shift measurement~\cite{2022_Ohayon} yields the $2P_{1/2} - 2P_{3/2}$ splitting:
$$
    \nu_\mathrm{FS,2P}^\M = 10918.2 \pm 7.4\,\mathrm{MHz},
$$
in agreement with QED.

\begin{table}[h!]
    \centering
    \begin{tabular}{c|cc}
                & Shift & Uncertainty \\
                & [$\mathrm{MHz}$] & [$\mathrm{MHz}$] \\
                \hline
        Fitting (stat.) & $-3.4$ & $6.0$\\
        Waveguide misalignment & & $3.2$\\
        RF field map & & $1.2$\\
        RF power stability & & $0.8$\\
        2\textsuperscript{nd}-order Doppler & & $0.6$\\   
        \hline\hline
        $2S_{1/2}-2P_{3/2}$ & \multicolumn{2}{l}{$9871.0 \pm 7.0$} \\
        Theory~\cite{2025_Blumer_FScalc} & \multicolumn{2}{l}{$9874.367 \pm 0.001$} \\
        \hline
        $2S_{1/2}-2P_{1/2}$~\cite{2022_Ohayon} & \multicolumn{2}{l}{$1047.2\pm 2.5$} \\
        Theory~\cite{2022_Janka_EXA} & \multicolumn{2}{l}{$1047.498 \pm 0.002$} \\
        \hline
        $2P_{1/2}-2P_{3/2}$ & \multicolumn{2}{l}{$10918.2 \pm 7.4$} \\
        Theory~\cite{2025_Blumer_FScalc} & \multicolumn{2}{l}{$10921.8639 \pm 0.0001$}
    \end{tabular}
    \caption{Summary of the \M\ fine structure measurement and derived $2P$ splitting. For completeness, the latest Lamb shift measurement and corresponding QED predictions are included.}
    \label{tab:M_results}
\end{table}

\section{Conclusion}
\label{sec:conclusion}

We have performed a precise microwave spectroscopy of the \M\ $2S_{1/2}-2P_{3/2}$ fine structure transition at PSI. Using a WR90 waveguide in the $10\,\mathrm{GHz}$ range, adapted from the Lamb shift apparatus, we validated the experimental setup and analysis through independent \Hy\ measurements, which agreed within $1.5\,\sigma$ of both theoretical predictions~\cite{2016_Horbatsch_Htabulation} and the most precise experimental benchmark~\cite{1994_Hagley_Hydrogen_FS}. 

The measured \M\ transition frequency,
$$
    \nu_\mathrm{FS}^\M = 9871.0 \pm 7.0~\mathrm{MHz},
$$
is consistent with the QED prediction~\cite{2025_Blumer_FScalc} and improves the previous result~\cite{1990_Kettell} by a factor of five. Combined with the recent Lamb shift measurement~\cite{2022_Ohayon}, this result yields a $2P_{1/2}-2P_{3/2}$ splitting of $10918.2 \pm 7.4\,\mathrm{MHz}$, improving the spectroscopic characterization of the $n=2$ manifold in \M\ and providing validation of bound-state QED in a purely leptonic two-body system.

The dominant systematic uncertainties, due to waveguide–beam alignment, RF standing-wave effects, and power stability, were quantified through detailed \textsc{Geant4} simulations. With the expected increase in beam intensity from the planned high-intensity muon source upgrade (HiMB) at PSI~\cite{2021_Aiba_HIMB}, in combination with the muCool scheme~\cite{2021_Antognini_muCool} and optimized microwave configurations, sub-MHz precision appears feasible \cite{2025_Blumer_FScalc}. Pushing the accuracy to the $\sim10\mathrm{kHz}$ level would enable sensitivity to higher-order recoil corrections and possible muon-specific interactions, thereby complementing other precision muon experiments such as the muon $g{-}2$ and muonic atom spectroscopy. The present work demonstrates the feasibility of precision \M\ spectroscopy with controlled systematics and establishes a foundation for future measurements testing bound-state QED and exploring possible new physics.

\begin{acknowledgments}
All measurements were performed at the Swiss Muon Source S$\mu$S, Paul Scherrer Institute, Villigen, Switzerland. This work was supported by the ERC consolidator grant 818053-Mu-MASS, the Swiss National Science Foundation under grants 197346, 219485, and 220823, and the ISF grant no. 2071390.
\end{acknowledgments}

\appendix
\section{Theory of the resonance lineshape}\label{app:Bloch}
The theoretical model presented here aims to describe the transition probabilities between quantum states when the atom is subjected to an oscillating electric field. This enables the determination of the resonance center from the measured spectral lines. Following the derivation of Ref.~\cite{2017_Marsman,2018_Marsman} the density operator $\hat{\rho}$ and the Hamiltonian $\hat{H}$ are used.

The time evolution of the quantum state is governed by the von Neumann equation for the density matrix $\hat{\rho}$:
\begin{equation}
\frac{\partial\hat{\rho}}{\partial t} = -\frac{i}{\hbar}[\hat{H},\hat{\rho}] = -\frac{i}{\hbar}\left[ \hat{H}_0 + \hat{D}\cdot\hat{V}, \hat{\rho} \right],
\end{equation}
where $\hat{H}_0$ is the time-independent diagonal Hamiltonian representing the energy eigenstates, and $\hat{H}_I = \hat{D}\cdot\hat{V}$ is the time-dependent interaction with the electric field. The dipole operator is $\hat{D} = -e\hat{r}$, and the field operator $\hat{V}$ describes an oscillating electric field perpendicular to the atomic beam.

For an $n$-level system, the density matrix $\hat{\rho}$ is an $n \times n$ matrix. The field interaction Hamiltonian introduces off-diagonal elements proportional to the electric dipole coupling strength, described by the Rabi frequencies $\Omega_{ij}$:
\begin{equation}
\label{eq:app_rabi}
\Omega_{ij} = -\frac{V_0}{\hbar} \bra{i} \hat{\epsilon} \cdot \hat{r} \ket{j} = -\frac{V_0}{\hbar} \cos(\omega t + \phi) \bra{i} \hat{r} \ket{j},
\end{equation}
where $V_0$ is the electric field amplitude, $\omega$ its angular frequency, and $\phi$ the phase. The dipole matrix elements $\bra{i} \hat{r} \ket{j}$ for transitions in the $n=2$ manifold are given in Table~\ref{tab:dipole_matrix} (see also Ref.~\cite{2021_Walraven}).

\begin{table}[htbp]
\centering
\begin{tabular}{c|ccc}
$nL_J, F, m_F$ & ($2S_{1/2},0,0$) & ($2S_{1/2}, 1, 0$) & ($2S_{1/2}, 1,\pm 1$) \\
\hline
$2P_{1/2}, 0, 0$ & $0$ & $-\sqrt{3}ea_0$ & $0$ \\
$2P_{1/2}, 1, 0$ & $-\sqrt{3}ea_0$ & $0$ & $0$ \\
$2P_{1/2}, 1, \pm1$ & $0$ & $0$ & $\mp\sqrt{3}ea_0$ \\
\hline
$2P_{3/2}, 1, 0$ & $\sqrt{6}ea_0$ & $0$ & $0$ \\
$2P_{3/2}, 1, \pm1$ & $0$ & $0$ & $\mp\sqrt{3/2}ea_0$ \\
$2P_{3/2}, 2, 0$ & $0$ & $\sqrt{6}ea_0$ & $0$ \\
$2P_{3/2}, 2, \pm1$ & $0$ & $0$ & $\pm 3/\sqrt{2}ea_0$ \\
$2P_{3/2}, 2, \pm2$ & $0$ & $0$ & $0$
\end{tabular}
\caption{Electric dipole matrix elements $\bra{i} \hat{r} \ket{j}$ for allowed transitions in the $n=2$ manifold of \Hy\ and \M~\cite{2021_Walraven}, where $a_0$ denotes the Bohr radius and $e$ the elementary electric charge.}
\label{tab:dipole_matrix}
\end{table}

Due to dipole selection rules, many transitions can be modeled using a two-level approximation, as was done in previous Lamb shift measurements with \M~\cite{2022_Ohayon,2022_Janka}. However, for initial states such as $2S_{1/2}, F=1, m_F=\pm1$, two final states $2P_{3/2}, F=1, m_F=\pm1$ and $2P_{3/2}, F=2, m_F=\pm1$ are accessible, necessitating a three-level treatment.

Using $\Omega_{ij} = \Omega_{ji}^*$, the time evolution of the system is described by the following set of coupled Bloch equations:\\
\begin{equation}
\label{eq:3levelBE}
    \begin{split}
    \dot{\rho}_{11} &= i\Omega_{12}(\rho_{12}-\rho_{21})+i\Omega_{13}(\rho_{13}-\rho_{31}) - \gamma_{2S}\rho_{11} \\
    \dot{\rho}_{22} &= -i\Omega_{12}(\rho_{12}-\rho_{21}) - (\gamma_{2S}+\gamma_{2P})\rho_{22} \\
    \dot{\rho}_{33} &= -i\Omega_{13}(\rho_{13}-\rho_{31}) - (\gamma_{2S}+\gamma_{2P})\rho_{33} \\
    \dot{\rho}_{12} &= \frac{-i(E_1 - E_2)}{\hbar}\rho_{12} -i\Omega_{12}(\rho_{22}-\rho_{11}) \\ &- i\Omega_{13}\rho_{23}^* - \frac{\gamma_{2S}+\gamma_{2P}}{2}\rho_{12}\\
    \dot{\rho}_{13} &= \frac{-i(E_1 - E_3)}{\hbar}\rho_{13} -i\Omega_{13}(\rho_{33}-\rho_{11}) \\ &- i\Omega_{12}\rho_{23} - \frac{\gamma_{2S}+\gamma_{2P}}{2}\rho_{13} \\
    \dot{\rho}_{23} &= \frac{-i(E_2 - E_3)}{\hbar}\rho_{23} -i\Omega_{12}\rho_{13} \\ &- i\Omega_{13}\rho_{12}^* - (\gamma_{2S}+\gamma_{2P})\rho_{23}
    \end{split}
\end{equation}
where the decay rates are $\gamma_i = 1/\tau_i$ for the lifetimes $\tau_i$ of the $2S$ and $2P$ states. These equations are numerically integrated to obtain the transition probabilities for a given RF exposure time $T$ and frequency $f$.

%\pagebreak
\bibliography{references}

\end{document}